\documentclass[conference]{IEEEtran}
\IEEEoverridecommandlockouts
\usepackage{cite}
\usepackage{amsmath,amssymb,amsfonts}
\usepackage{algorithmic}
\usepackage{graphicx}
\usepackage{textcomp}
\usepackage{epsfig}
\usepackage{multirow}
\usepackage{steinmetz}
\usepackage{float}
\usepackage{subfig}
\newcommand{\Ntx}{N_{\mathrm{tx}}}
\newcommand{\bhx}{\mathbf{h}_\mathrm{x}}
\newcommand{\bhy}{\mathbf{h}_\mathrm{y}}
\newcommand{\bfx}{\mathbf{f}_\mathrm{x}}
\newcommand{\bfy}{\mathbf{f}_\mathrm{y}}
\newcommand{\Ptx}{P_\mathrm{tx}}
\newcommand{\fxk}{f_{\mathrm{x},k}}
\newcommand{\gbx}{{\gamma}_{\mathrm{b},\mathrm{x}}}
\newcommand{\gby}{{\gamma}_{\mathrm{b},\mathrm{y}}}
\newcommand{\hxk}{h_{\mathrm{x},k}}
\newcommand{\hyk}{h_{\mathrm{y},k}}
\newcommand{\fyk}{f_{\mathrm{y},k}}
\newcommand{\Lb}{\lambda}
\newcommand{\imj}{\mathsf{j}}
\newcommand{\Ldp}{\ell_{\mathrm{dp}}}
\usepackage{booktabs}
\def\BibTeX{{\rm B\kern-.05em{\sc i\kern-.025em b}\kern-.08em
		T\kern-.1667em\lower.7ex\hbox{E}\kern-.125emX}}
\begin{document}
	
	\title{Near-field focusing using phased arrays with dynamic polarization control\\
	}
	
 \author{Nitin Jonathan Myers$^{\ast}$, Yanki Aslan$^{\dagger}$ and Geethu Joseph$^{\dagger}$ \\
		$^{\ast}$Delft Center for Systems and Control, Delft University of Technology, The Netherlands\\
			$^{\dagger}$Department of Microelectronics, Delft University of Technology, The Netherlands\\
		 Email: $\{$N.J.Myers,Y.Aslan, G.Joseph$\}$@tudelft.nl}
	\maketitle

\begin{abstract} 
Phased arrays in near-field communication allow the transmitter to focus wireless signals in a small region around the receiver.  Proper focusing is achieved by carefully tuning the phase shifts and the polarization of the signals transmitted from the phased array.  In this paper, we study the impact of polarization on near-field focusing and investigate the use of dynamic polarization control (DPC) phased arrays in this context.  Our studies indicate that the optimal polarization configuration for near-field focusing varies spatially across the antenna array.  Such a spatial variation motivates the need for DPC phased arrays which allow independent polarization control across different antennas.  We show using simulations that DPC phased arrays in the near-field achieve a higher received signal-to-noise ratio than conventional switched- or dual-polarization phased arrays.
\end{abstract}
\vspace{-3mm}
\section{Introduction}
\par Near-field wireless systems are those in which the distance between the transmitter (TX) and the receiver (RX) is less than the Fraunhofer distance~\cite{NF_fraunhoffer}. As the Fraunhofer distance is smaller at higher carrier frequencies, such systems are increasingly important in sub-THz and THz systems.  Some applications where near-field communication is promising include kiosk download stations, data centers, and wearables~\cite{THz_applications}.  The use of phased arrays for near-field communication enables focused transmission of radio frequency (RF) signals. Focusing is usually achieved by phase shifting the RF signals transmitted across the array and can be interpreted as the near-field counterpart of the common directional beamforming~\cite{NF_focus}.
\par In mobile near-field applications,  the received signal power can decrease substantially when the RX rotates around its axis.  The power reduction is a consequence of polarization mismatch and occurs even when the RF signals are focused at the RX. Common methods to mitigate polarization mismatch are based on switched-polarization beamforming~\cite{sw_dual_pol} or dual-polarization beamforming~\cite{dual_pol_BF2}. These architectures allow control over the polarization of the RF signals transmitted from an array.  A DPC phased array architecture was proposed in~\cite{DPC_hw2} to tune the polarization of the RF signal transmitted from every antenna, thereby providing greater flexibility than conventional switched- or dual-polarization arrays.  
\par In this paper, we consider a near-field line-of-sight (LoS) communication scenario and investigate the use of DPC phased arrays from a signal processing perspective.  We observe that the optimal polarization configuration varies spatially across the antenna array, unlike far-field LoS scenarios where the optimal configuration is spatially invariant.  As a result, standard phased arrays based on switched- or dual linear-polarization (dubbed dual-polarization) control, which are well suited to far-field systems, perform poor in near-field scenarios.  In this paper, we show that DPC phased arrays result in a higher received SNR over switched- and dual-polarization phased arrays at short distances. The SNR improvement achieved with DPC phased arrays over the dual-polarization counterpart, however, decreases with the transceiver distance and becomes negligible in the far-field limit. 
\par Most of the prior work on near-field LoS communication assumes perfect polarization alignment between the TX and the RX~\cite{NF_rotULA, NF_chest,NF_chest2,InFocus,WB_BF}. For instance, the channel capacity of near-field LoS systems was studied in~\cite{NF_rotULA} under mechanical antenna steering. Near-field LoS channel estimation techniques were developed in~\cite{NF_chest,NF_chest2}, and broadband beamforming robust to the misfocus effect was investigated in~\cite{InFocus,WB_BF}.  The perfect polarization alignment assumption in~\cite{NF_rotULA, NF_chest,NF_chest2,InFocus,WB_BF} is unrealistic, especially in near-field scenarios, as any mismatch in the orientation of the arrays can degrade the received signal power.  Prior work has also developed new antenna architectures to mitigate polarization mismatch in near-field systems.  For instance,  joint beamforming and polarization tuning was demonstrated in~\cite{JBFPT} for a far-field setup using an intelligent reflecting surface.  New antenna designs were proposed in~\cite{HW_pol1,HW_pol2,HW_pol3} to mitigate polarization mismatch in the near-field.  The designs in ~\cite{HW_pol1,HW_pol2,HW_pol3} assume a particular orientation of the RX and are fixed. This paper investigates the use of DPC phased arrays which can adapt the polarization configuration according to the orientation of the RX.
\par Notation: $a$ denotes a complex scalar with magnitude $|a|$ and phase $\phase{a}$.  $a^{\ast}$ is the complex conjugate of $a$. We use $\hat{\mathbf{x}}$, $\hat{\mathbf{y}}$, and $\hat{\mathbf{z}}$ to represent the unit vectors along the $x$, $y$, and $z$ axes.  A vector $\overrightarrow{\mathbf{a}}$ in 3D-space is expressed as $\overrightarrow{\mathbf{a}}=a_x\hat{\mathbf{x}}+a_y\hat{\mathbf{y}}+a_z\hat{\mathbf{z}}$.  For vectors $\overrightarrow{\mathbf{a}}$ and $\overrightarrow{\mathbf{b}}$,  we use $\overrightarrow{\mathbf{a}} \mathbf{\cdot} \overrightarrow{\mathbf{b}}$ to denote their dot product and $\overrightarrow{\mathbf{a}} \times \overrightarrow{\mathbf{b}}$ to denote their cross product. The length of $\overrightarrow{\mathbf{a}}$ is $\Vert \overrightarrow{\mathbf{a}} \Vert=\sqrt{a^2_x + a^2_y + a^2_z}$. Finally, $\mathsf{j}=\sqrt{-1}$.
\section{System and channel model}
\par We consider a circular uniform planar array at the TX with $\Ntx$ antennas on the $xy$ plane, bounded within a circle of radius $R$.  Although we assume a circular array for symmetry, our solution naturally extends to rectangular array configurations. Every antenna at the TX is enabled with dynamic polarization control. A simple way to implement DPC is to use a pair of dipoles mounted at $90^{\circ}$ at each antenna.  Then, separate amplitude and phase control is applied over the RF signal fed to each dipole as shown in Fig. 1. This allows configuring the polarization of the electric field radiated by an antenna. For example, linearly polarized electric field can be radiated by applying the same phase across both the dipoles; the polarization angle can be tuned by changing the amplitude control across the dipoles.  Similarly, circular polarization can be achieved by applying the same amplitude and $90^{\circ}$ phase shift across the dipoles.  We assume that the length of each TX dipole is $\Ldp=\Lb/2$, where $\Lb$ is the carrier wavelength, and that the dipoles are oriented along the $x$- and the $y$- axis.
\begin{figure}[h]
\centering
\includegraphics[trim=0cm 0cm 0cm 0.4cm, width=0.3\textwidth]{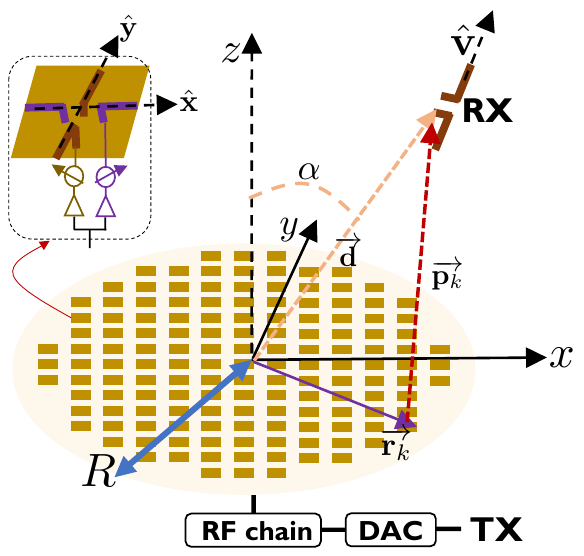}
\caption{\small A near-field LoS communication scenario where the TX is equipped with a DPC phased array of $\Ntx$ antennas and the RX has a single dipole antenna.  Each of the $\Ntx$ antennas at the TX comprises two dipoles,  two amplifiers, and two phase shifters.}
\normalsize
  \label{fig:TX_RX}
\end{figure}
\par We consider a single dipole antenna of length $\Ldp$ at the RX, which is a reasonable choice for low-power and low-cost radios~\cite{IOT_dipole}. We use $\hat{\mathbf{v}}$ to denote the unit vector along the direction of the receive dipole. By the ``symmetric'' nature of the TX array, we assume that the RX lies on the $xz$ plane without loss of generality. The center of the TX array is defined to be $(0,0,0)$ and the position vector of the RX center is defined as $\overrightarrow{\mathbf{d}}$.  The angle made by $\overrightarrow{\mathbf{d}}$ with the $z$-axis is denoted by $\alpha$.  The transceiver distance is defined as $d=\Vert\overrightarrow{\mathbf{d}} \Vert$. Further, we define $\bhx \in \mathbb{C}^{\Ntx}$ as the narrowband channel between the TX dipoles oriented along the $x$-axis and the receiver. The $\Ntx$ entries in $\bhx$ are arranged according to a certain ordering of the antennas. For the same ordering, the channel between the TX dipoles along the $y$-axis and the receiver is denoted by $\bhy \in \mathbb{C}^{\Ntx}$.  The TX applies antenna weight vectors $\bfx  \in \mathbb{C}^{\Ntx}$ and $\bfy  \in \mathbb{C}^{\Ntx}$ to its dipoles oriented along $x$- and $y$- directions. The antenna weights are realized using amplifiers and phase shifters.  Under the narrowband assumption, the received signal when the TX transmits a unit norm symbol $s$ is 
\begin{equation}
\label{eq:sys_model}
y=(\bhx^T\bfx + \bhy^T\bfy)\sqrt{\Ptx}s +v,
\end{equation}
where $\Ptx$ is the transmit power and $v$ is additive white Gaussian noise with variance $\sigma^2$.  We make the narrowband assumption for simplicity and also to ignore the misfocus effect which arises in high bandwidth near-field systems~\cite{InFocus}. 
\par
\begin{figure}[h]
\centering
\includegraphics[trim=0cm 0.1cm 0cm 0.35cm, width=0.24\textwidth]{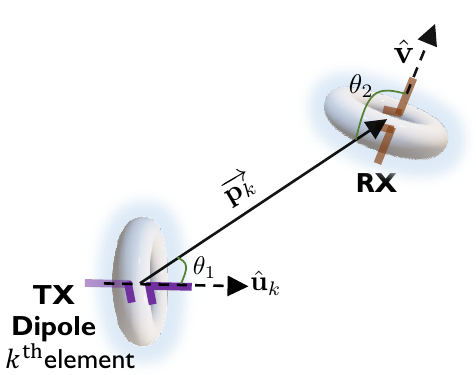}
\caption{\small A near-field LoS communication scenario showing one TX dipole antenna and the RX dipole.  The TX and the RX dipoles are oriented along  $\hat{\mathbf{u}}_k$ and  $\hat{\mathbf{v}}$. The polarized channel depends on the field pattern and the orientation of the dipoles.}
\normalsize
  \label{fig:Pair_Dipoles}
\end{figure}

 We now describe the polarized channel between a particular TX dipole element and the RX dipole.  We use $\hat{\mathbf{u}}_k$ to denote the unit vector along the direction of the $k^{\mathrm{th}}$ TX dipole. The position vector of the RX relative to the TX is defined as $\overrightarrow{\mathbf{p}_k}$, and a unit vector along this direction is $\hat{\mathbf{p}}_k$.  We observe from Fig. \ref{fig:TX_RX} that $\overrightarrow{\mathbf{p}_k}=\overrightarrow{\mathbf{d}}-\overrightarrow{\mathbf{r}_k}$,  where $\overrightarrow{\mathbf{r}_k}$ is the position vector of the $k^{\mathrm{th}}$ TX dipole.  A sketch of these vectors is shown in Fig. \ref{fig:Pair_Dipoles}.  The polarized channel between the TX and the RX comprises three terms: a) the unpolarized channel due to free-space path loss and propagation delay, b) the normalized field patterns of the transmitting and receiving antennas, and c) the inner product between the receive dipole direction and the impinging electric field.  All these terms vary spatially in near-field communication.  As a result, the composite of these three effects must be computed between every TX antenna and the RX antenna to obtain the channel.  
\par The unpolarized channel between the $k^{\mathrm{th}}$ TX dipole and the RX dipole shown in Fig. \ref{fig:Pair_Dipoles} is defined by
\begin{equation}
\label{eq:h_up}
h^{\mathrm{up}}_k=\frac{\Lb}{4 \pi \Vert \overrightarrow{\mathbf{p}_k} \Vert} \mathrm{exp}\left(-\imj 2 \pi  \Vert \overrightarrow{\mathbf{p}_k} \Vert/{\Lb}\right).
\end{equation}
We define $\theta_{\mathrm{tx},k}$ as the angle between $\hat{\mathbf{u}}_k$ and $\hat{\mathbf{p}}_k$, and $\theta_{\mathrm{rx},k}$ as the angle between $\hat{\mathbf{v}}$ and $-\hat{\mathbf{p}}_k$.  Mathematically,  $\theta_{\mathrm{tx},k}= \mathrm{cos}^{-1}(\hat{\mathbf{u}}_k \mathbf{\cdot} \hat{\mathbf{p}}_k)$ and $\theta_{\mathrm{rx},k}=\pi - \mathrm{cos}^{-1}(\hat{\mathbf{v}} \mathbf{\cdot} \hat{\mathbf{p}}_k)$. The normalized field pattern at the TX dipole in the direction of $\hat{\mathbf{p}}_k$ is~\cite{Ant_text} 
\begin{equation}
\label{eq:f_tx}
g_{\mathrm{tx},k}= \left[\mathrm{cos}(\pi \Ldp\mathrm{cos} \theta_{\mathrm{tx},k}/ \Lb)-\mathrm{cos}(\pi \Ldp/ \Lb)\right]/\mathrm{sin}\theta_{\mathrm{tx},k}.
\end{equation}
Similarly, the normalized field pattern at the RX dipole $g_{\mathrm{rx},k}$ along $\hat{\mathbf{p}}_k$ is obtained by using $\theta_{\mathrm{rx},k}$ instead of $\theta_{\mathrm{tx},k}$ in \eqref{eq:f_tx}.  The unit vector along the electric field impinging at the RX due to the $k^{\mathrm{th}}$ TX dipole element is~\cite{Ant_text}
\begin{equation}
\label{eq:imping_field}
\hat{\mathbf{e}}_k=\frac{\hat{\mathbf{p}}_k \times ( \hat{\mathbf{u}}_k \times \hat{\mathbf{p}}_k)}{\Vert \hat{\mathbf{p}}_k \times ( \hat{\mathbf{u}}_k \times \hat{\mathbf{p}}_k) \Vert}.
\end{equation}
The difference between the receive dipole direction $ \hat{\mathbf{v}}$ and the impinging electric field $\hat{\mathbf{e}}$ creates polarization loss. We define ${\beta}_{\mathrm{pol},k}=\hat{\mathbf{v}} \mathbf{\cdot} \hat{\mathbf{e}}_k$. The polarized channel between the $k^{\mathrm{th}}$ TX dipole oriented along $\hat{\mathbf{u}}_k$ and the RX is then~\cite{MassMIMO}
\begin{equation}
\label{eq:eff_channel}
h_{\mathrm{pol}, k}=h^{\mathrm{up}}_k g_{\mathrm{tx},k} g_{\mathrm{rx},k} {\beta}_{\mathrm{pol},k}.
\end{equation}
The polarized channel $\bhx$ is computed between every TX dipole oriented along the $x$ axis, i.e., with $\hat{\mathbf{u}}_k=\hat{\mathbf{x}}$, and the RX using \eqref{eq:eff_channel}. In the same way,  the entries of $\bhy$ are computed from \eqref{eq:eff_channel} by setting $\hat{\mathbf{u}}_k=\hat{\mathbf{y}}$.
\section{Polarization-aware near-field focusing}
In this section, we discuss an optimization problem for beamforming (focusing) with DPC, and study the polarization associated with the optimized beamformer.  Finally, we compare DPC-based focusing with other benchmarks. 
\subsection{Focusing with DPC}
We assume the common per-antenna power constraint at every DPC antenna, i.e.,
\begin{equation}
\label{eq:pow_const}
|f_{\mathrm{x},k}|^2+|f_{\mathrm{y},k}|^2=1/\Ntx\,\, \forall k.
\end{equation}
From \eqref{eq:sys_model}, we observe that the received SNR is $\Ptx | \bhx^T \bfx + \bhy ^T \bfy|^2/\sigma^2$. The beamformer that maximizes SNR is then
\begin{equation}
\label{eq:BF_optim}
\{\bfx^{\mathrm{opt}},\bfy^{\mathrm{opt}}\}= \begin{array}{c}
\underset{\bfx, \bfy}{\mathrm{argmax}} \, | \bhx^T \bfx + \bhy ^{T} \bfy|\\
\mathrm{subject}\,\mathrm{to}\,\,  |f_{\mathrm{x},k}|^2+|f_{\mathrm{y},k}|^2=1/\Ntx \,\, \forall k.\\
\end{array}
\end{equation}
The optimization problem in \eqref{eq:BF_optim} accounts for the unpolarized and the polarized components of the channel.
\par We simplify the objective in \eqref{eq:BF_optim} to obtain a closed form solution for $\bfx^{\mathrm{opt}}$ and $\bfy^{\mathrm{opt}}$.  Specifically,
\begin{align}
\nonumber
|\bhx^T \bfx + \bhy ^{T} \bfy| &= \big|\sum^{\Ntx}_{k=1} \left(\fxk \hxk +\fyk \hyk \right) \big|\\
\nonumber
&= \big|\sum^{\Ntx}_{k=1} |\fxk||\hxk|\mathrm{exp}\{ \imj(\phase{\hxk}+\phase{\fxk}) \} \\
\label{eq:expand_sum}
&+ \sum^{\Ntx}_{k=1}|\fyk||\hyk|\mathrm{exp}\{\imj(\phase{\hyk}+\phase{\fyk}) \} \big|
\end{align}
The right hand side of \eqref{eq:expand_sum}, comprising a sum of $2\Ntx$ complex numbers,  is maximized when all the complex numbers in the summation align on top of each other.  All the $2\Ntx$ complex numbers align when 
\begin{equation}
\label{eq:optimized_phase}
\phase{\fxk^{\mathrm{opt}}}=-\phase{\hxk}\;\; \mathrm{and}\;\; \phase{\fyk^{\mathrm{opt}}}=-\phase{\hyk} \, \forall k.
\end{equation}
The optimized phase values are same as that of $\bhx^{\ast}$ and $\bhy^{\ast}$.
\par Now, we determine the magnitudes $|\fxk^{\mathrm{opt}}|$ and $|\fyk^{\mathrm{opt}}|$ of the optimized beamformer.  With the optimized phase values of $\fxk^{\mathrm{opt}}$ and $\fxk^{\mathrm{opt}}$, the objective in \eqref{eq:BF_optim} simplifies to $\sum^{\Ntx}_{k=1} |\fxk||\hxk| +|\fyk||\hyk|$. As the constraints in \eqref{eq:BF_optim} are defined independently for each $k$, the optimization problem for the magnitudes can be decoupled as
\begin{equation}
\label{eq:decoup_max}
\{|\fxk^{\mathrm{opt}}|, \,|\fyk^{\mathrm{opt}}|\}=\begin{array}{c}
\underset{|\fxk|, |\fyk|}{\mathrm{argmax}} \, |\fxk||\hxk| +|\fyk||\hyk|\\
\mathrm{subject}\,\mathrm{to}\,\,  |f_{\mathrm{x},k}|^2+|f_{\mathrm{y},k}|^2=1/\Ntx\\
\end{array},
\end{equation}
for each $k$. The objective in \eqref{eq:decoup_max} can be interpreted as the inner product between $(|\fxk|,|\fyk|)$ and $(|\hxk|,|\hyk|)$, which achieves its maximum when the vectors are aligned under the power constraint, i.e.,
\begin{align}
\nonumber
|\fxk^{\mathrm{opt}}|&=\frac{|\hxk|}{\sqrt{\Ntx(|\hxk|^2+|\hyk|^2)}}\;\; \mathrm{and}\\
\label{eq:optimized_mag}
|\fyk^{\mathrm{opt}}|&=\frac{|\hyk|}{\sqrt{\Ntx(|\hxk|^2+|\hyk|^2)}}\, \forall k.
\end{align}
We observe from \eqref{eq:optimized_phase} and \eqref{eq:optimized_mag} that $\bfx^{\mathrm{opt}}$ and $\bfy^{\mathrm{opt}}$ are antenna-wise normalized versions of the conjugate beamformers $\bhx^{\ast}$ and $\bhy^{\ast}$, respectively. 
\subsection{Polarization distribution of the transmitted electric field}
With the DPC architecture shown in Fig. \ref{fig:TX_RX}, the same current is amplified and phase shifted differently before it is applied to the dipoles.  At the $k^{\mathrm{th}}$ antenna, the electric field generated at the $x$-oriented dipole is $E_\mathrm{o} \fxk^\mathrm{opt} e^{\imj \omega_\mathrm{o} t}\hat{\mathbf{x}}$, where $E_\mathrm{o}$ is defined as the field generated due to a unit antenna weight and $\omega_\mathrm{o}$ is the carrier frequency. Similarly, $E_\mathrm{o} \fyk^{\mathrm{opt}} e^{\imj \omega_\mathrm{o} t}\hat{\mathbf{y}}$ is generated by the $y$-oriented dipole at the $k^{\mathrm{th}}$ antenna.
\par We now describe the polarization of the electric field generated at the $k^{\mathrm{th}}$ antenna. At this antenna, the electric field phasors associated with the $x$ and $y$-oriented dipoles are $E_\mathrm{o} |\fxk^{\mathrm{opt}}|\mathrm{exp}\left(\imj \phase{\fxk^{\mathrm{opt}}}\right)$ and $E_\mathrm{o} |\fyk^{\mathrm{opt}}|\mathrm{exp}\left(\imj \phase{\fyk^{\mathrm{opt}}}\right)$.  Equivalently, these phasors can also be expressed as $E_\mathrm{o} |\fxk^{\mathrm{opt}}|\mathrm{exp}\left(-\imj \phase{\hxk}\right)$ and $E_\mathrm{o} |\fyk^{\mathrm{opt}}|\mathrm{exp}\left(-\imj \phase{\hyk} \right)$ using \eqref{eq:optimized_phase}. In the complex plane, the angle between the two phasors is either $0^{\circ}$ or $180^{\circ}$ for the channel model in \eqref{eq:eff_channel}. This is because $h_k^{\mathrm{up}}$ is same across both the polarizations, and the remaining quantities in \eqref{eq:eff_channel}, i.e., $\{g_{\mathrm{tx},k},  g_{\mathrm{rx},k}$ and $g_{\mathrm{pol},k}\}$, are purely real.  As the electric fields generated along the $x$- and the $y$- directions only oscillate with either the same phase or the opposite phase, it follows that the $k^{\mathrm{th}}$ antenna transmits a linearly polarized wave.  Furthermore, the polarization angle of this wave is $\mathrm{tan}^{-1}(\fyk^{\mathrm{opt}}/\fxk^{\mathrm{opt}})$. 
\vspace{-6mm}
\begin{figure}[h!]
\centering
\subfloat[$d=15\,\mathrm{cm}$ and $\alpha=30^{\circ}$]{\includegraphics[trim=0.5cm 6.5cm 2cm 7cm,clip=true,width=4.5cm, height=4.7cm]{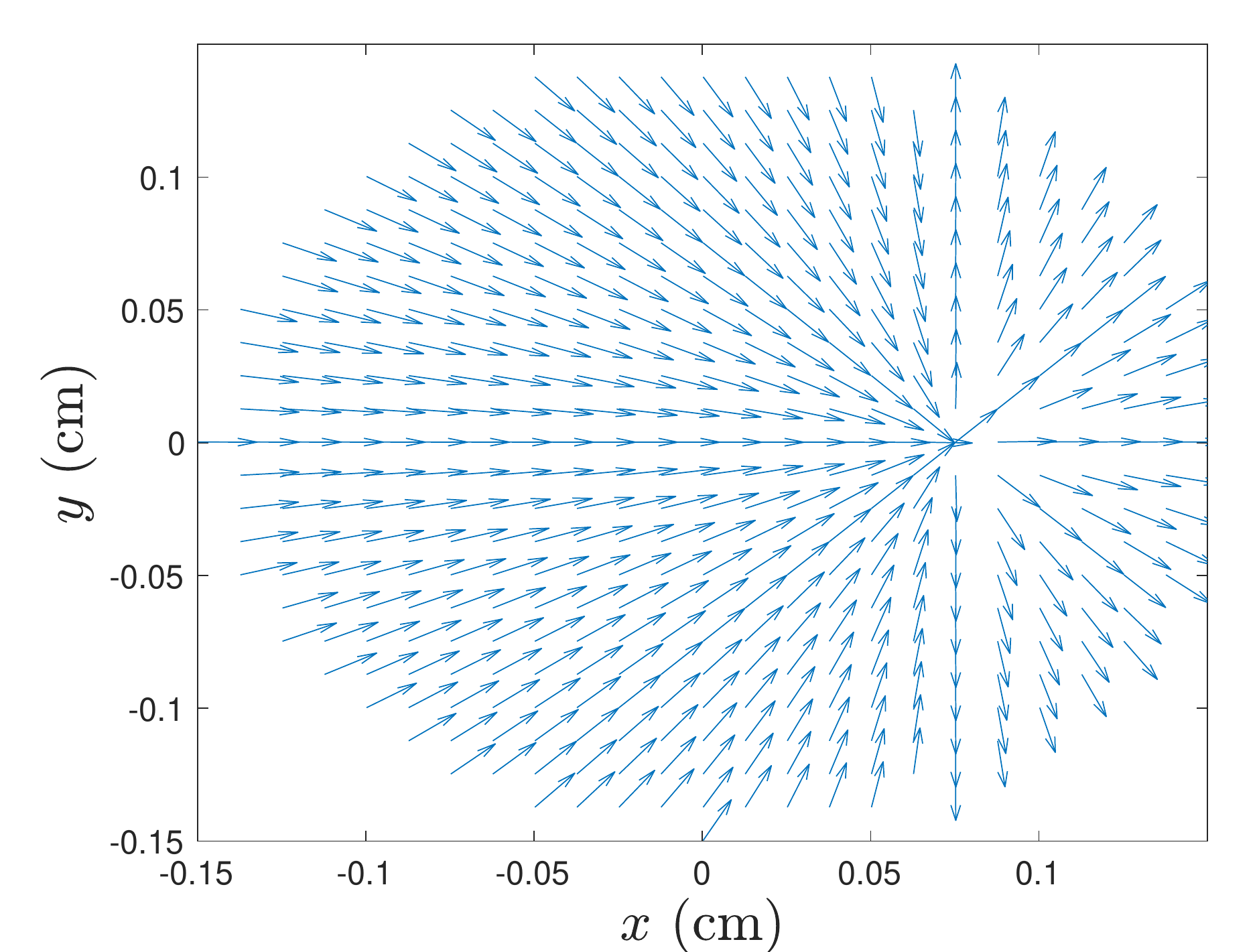}\label{fig:Pol_NF}} 
\subfloat[$d=100\,\mathrm{cm}$ and $\alpha=30^{\circ}$]{\includegraphics[trim=0.5cm 6.5cm 2cm 7cm,clip=true,width=4.5cm, height=4.7cm]{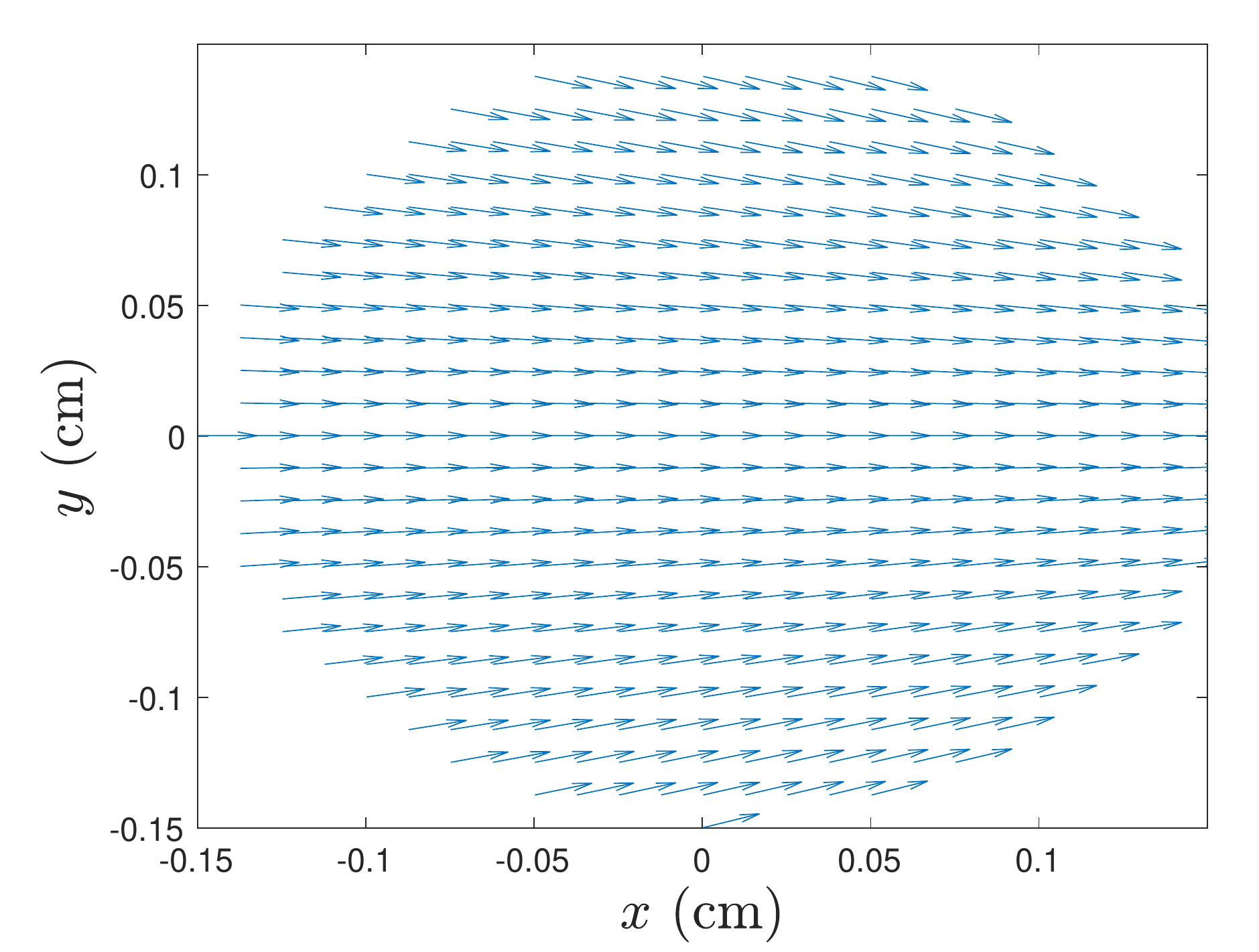}\label{fig:Pol_FF}}
\caption{\small For the optimized beamformer, the figure shows the polarization of the electric field generated at some of the antennas within the TX array of radius $R=15\,\mathrm{cm}$. The optimal polarization configuration varies spatially for $d=15\, \mathrm{cm}$, which motivates the need for DPC in near-field settings.
\normalsize}
\end{figure}
\par We study the polarization associated with the optimized TX beamformer, for the near-field scenario in Fig. \ref{fig:TX_RX}, for $\alpha=\pi/6$ and $\hat{\mathbf{v}}=\hat{\mathbf{z}}$.  We consider a TX array of radius $R=15\,\mathrm{cm}$ and a carrier frequency of $300\,\mathrm{GHz}$. The spatial polarization distribution within the TX is shown in Fig. \ref{fig:Pol_NF} and Fig. \ref{fig:Pol_FF}, for $d=15\, \mathrm{cm}$ and $d=100\,\mathrm{cm}$.  From Fig. \ref{fig:Pol_NF}, we notice that the polarization angle of the electric field, associated with the optimized beamformer, varies across the transmit antennas for $d=15\,\mathrm{cm}$.  The use of DPC phased arrays allows spatial tuning of the polarization angle to match the optimal configuration.  We observe from Fig. \ref{fig:Pol_FF} that the polarization angle is almost spatially invariant when the transceiver distance increases to $d=100\,\mathrm{cm}$. In such scenarios, standard dual-polarization arrays that can achieve a uniform polarization configuration across the array suffice.
\subsection{Benchmark architectures}
We compare DPC phased array-based beamforming with those achieved by switched- and dual-polarization phased arrays.  On the one hand, the switched-polarization architecture achieves beamforming using either the $x$-dipoles or the $y$- dipoles.  On the other hand, the dual-polarization architecture leverages all the dipoles by first performing digital beamforming (DIG) followed by phase shifting.  A schematic of these two benchmark architectures is shown in Fig.  4 for $\Ntx=2$.
\vspace{-3mm}
\begin{figure}[h!]
\centering
\subfloat[Dual-polarization]{\includegraphics[trim=0.1cm 0.3cm 2cm 0.7cm,clip=true,width=3cm, height=3cm]{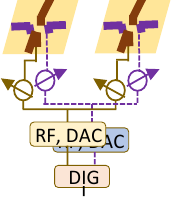}\label{fig:Dual_pol}} \:\:\:\:
\subfloat[Switched-polarization]{\includegraphics[trim=2cm 0.3cm 0.1cm 0.7cm,clip=true,width=3cm, height=3cm]{benchmarks.pdf}\label{fig:SW_pol}}
\caption{\small An illustration of dual-polarization and switched-polarization phased arrays with two antennas, each comprising two dipoles.  
\normalsize}
\end{figure}
\par We now describe the beamformers used in the benchmark dual-polarization and the switched-polarization architectures. In both the hardwares, the phase shifters linked to the $x$-dipoles are fed with antenna weights $\fxk^{\mathrm{b}}=\mathrm{exp}(-\imj \phase{\hxk})/\sqrt{\Ntx} \, \forall k$. Similarly, $\fyk^{\mathrm{b}}=\mathrm{exp}(-\imj \phase{\hyk})/\sqrt{\Ntx} \, \forall k$ is applied to the $y$-dipoles. We define $\gbx=|\bhx^T \bfx^{\mathrm{b}}|$ and $\gby=|\bhy^T \bfy^{\mathrm{b}}|$. Then, the received SNR with the switched-polarization architecture is $\mathrm{SNR}_{\mathrm{sw}}=\Ptx\mathrm{max}\{{\gbx^2}, { \gby^2} \}/{\sigma}^2$ and that with the dual-polarization architecture is $\mathrm{SNR}_{\mathrm{dp}}=\Ptx (\gbx^2+\gby^2)/{\sigma}^2$.  
\par Our paper ignores the differences in power consumption and insertion loss across the three architectures.  For example, the DPC phased array has additional amplitude control elements at each antenna when compared to the benchmarks.  The dual-polarization array has an additional DAC and RF chain, while the switched-polarization array has a switch.  A fair comparison across these architectures depends on the circuit technology and the resolution of the components within these arrays. 
\section{Simulation results}
We consider a wireless system operating in a near-field LoS scenario shown in Fig. \ref{fig:TX_RX}.  The TX array is of radius $R=15\,\mathrm{cm}$ and comprises half-wavelength spaced antenna elements.  The carrier frequency is $300\, \mathrm{GHz}$. Each TX antenna has two dipoles placed at $90^{\circ}$ to achieve DPC, while the RX has a single dipole.  The transceiver distance is set within the range $d\in [10\, \mathrm{cm}, 100\, \mathrm{cm}]$ and the operating bandwidth is chosen as $B=100\, \mathrm{MHz}$.  The delay spread of the near-field LoS channel in a boresight scenario is $(\sqrt{d^2+R^2}-d)/c$, where $c$ is the speed of light.  For the range of transceiver distances considered, the maximum of this delay spread evaluates to $0.26\, \mathrm{ns}\ll 1/B$.  Therefore, the narrowband assumption is reasonable and the misfocus effect can also be ignored~\cite{InFocus}.
\par We note that DPC-based beamforming results in an SNR of $\mathrm{SNR}_{\mathrm{DPC}}=\Ptx | \bhx^T \bfx^{\mathrm{opt}} + \bhy ^T \bfy^{\mathrm{opt}}|^2/\sigma^2$. The DPC phased array achieves an SNR that is never smaller than obtained with the benchmark hardwares. This is because the switched-polarization architecture can be interpreted as a special case of the DPC array, by setting the amplitudes associated with the dipoles oriented along a particular direction to zero. Similarly, the dual-polarization architecture is a special instance of the DPC array,  when the same amplitude is applied to the dipoles oriented along a particular direction, i.e., $\fxk=f_{\mathrm{x}}\, \forall k$ and $\fxk=f_{\mathrm{y}}\, \forall k$. Therefore, we evaluate the improvement in the received SNR due to the use of DPC phased array over the benchmarks, i.e., $10\mathrm{log}_{10}(\mathrm{SNR}_{\mathrm{DPC}}/\mathrm{SNR}_{\mathrm{sw}})$ and $10\mathrm{log}_{10}(\mathrm{SNR}_{\mathrm{DPC}}/\mathrm{SNR}_{\mathrm{dp}})$.
\vspace{-6mm}
\begin{figure}[h!]
\centering
\subfloat[$10\mathrm{log}_{10}(\mathrm{SNR}_{\mathrm{DPC}}/\mathrm{SNR}_{\mathrm{sw}})$]{\includegraphics[trim=2cm 6.5cm 2.5cm 7.5cm,clip=true,width=4.45cm, height=4.6cm]{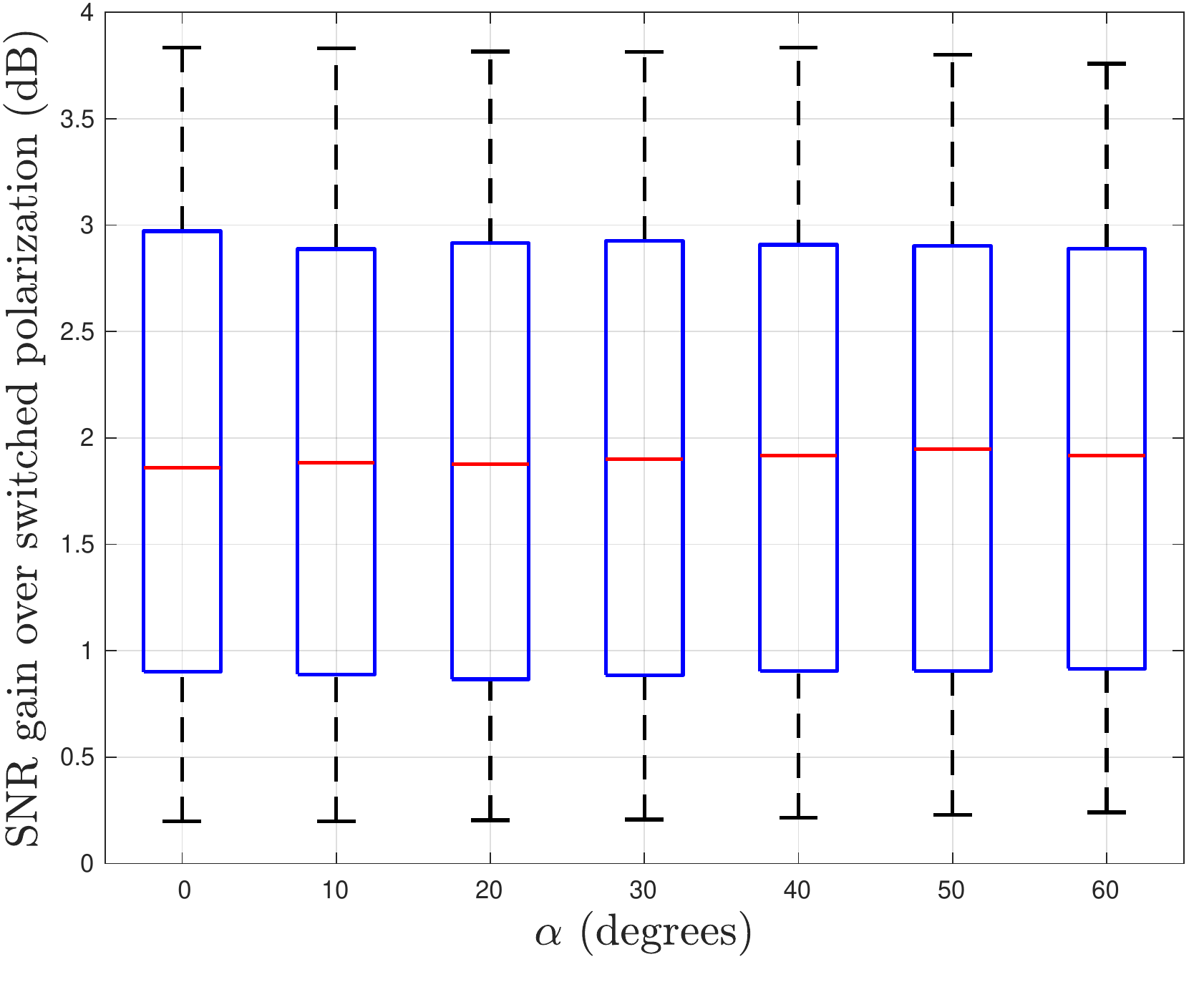}\label{fig:SNRimp_SW}} 
\subfloat[$10\mathrm{log}_{10}(\mathrm{SNR}_{\mathrm{DPC}}/\mathrm{SNR}_{\mathrm{dp}})$]{\includegraphics[trim=2cm 6.5cm 2.5cm 7.5cm,clip=true,width=4.45cm, height=4.6cm]{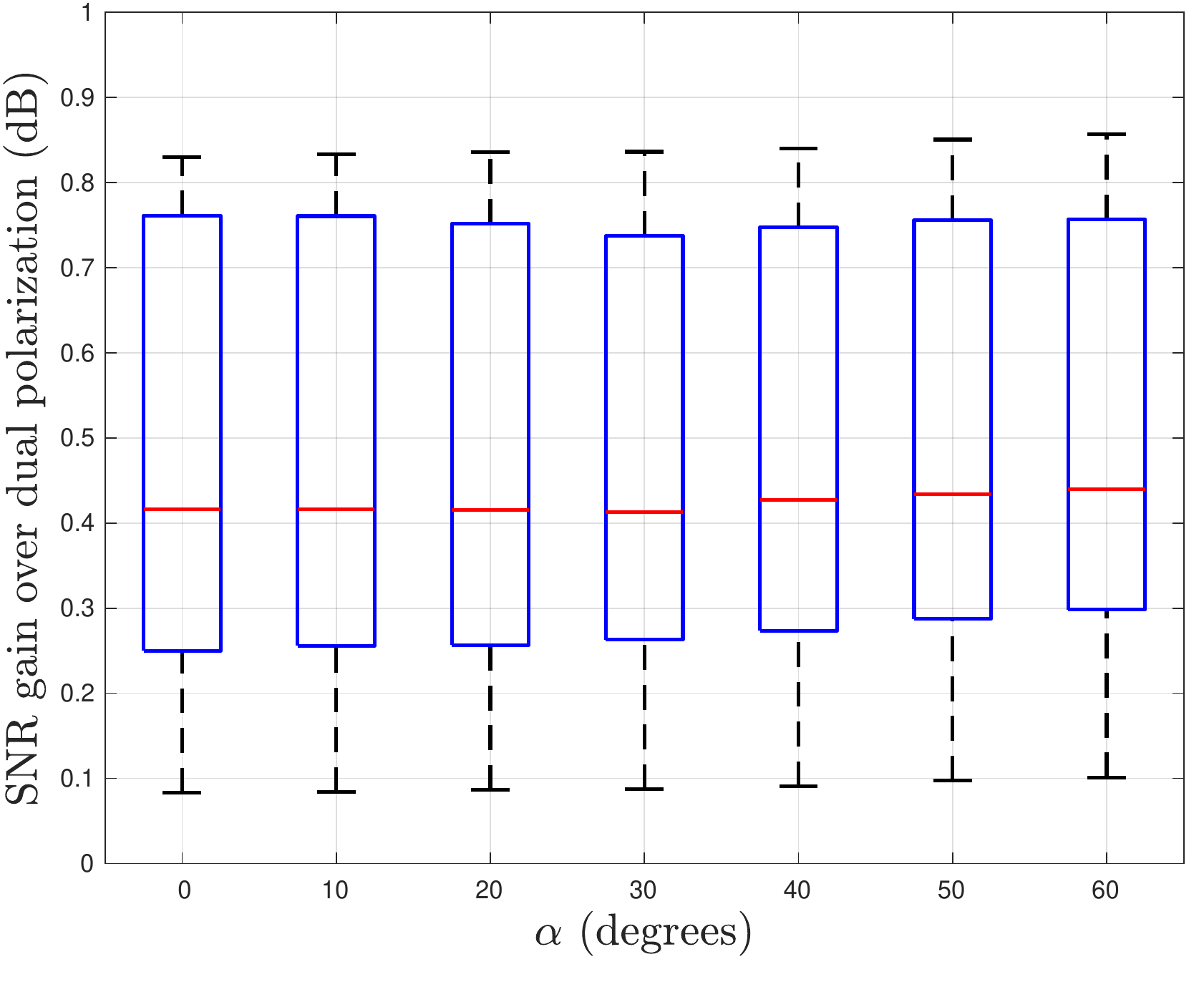}\label{fig:SNRimp_Dual}}
\caption{\small Improvement in the received SNR with the use of a DPC phased array over switched- or dual- polarization architectures.  The SNR improvement is shown as a function of $\alpha$ for $d=10\, \mathrm{cm}$.
\normalsize}
\end{figure}
\par We first study the SNR improvement with DPC as a function of the angle $\alpha$,  when the transceiver distance is $d=10\, \mathrm{cm}$. Note that the channel also depends on the orientation of the RX dipole.  The dipole is rotated about its axis in steps of $10^{\circ}$ along azimuth and elevation, resulting in different $\hat{\mathbf{v}}$ for each rotation. As a result, the rotation allows generating $36\times 18=648$ channel realizations for a particular $\alpha$ and $d$.  A distribution of received SNR is constructed with these $648$ channels, for a specific $\alpha$ and $d$. Then, a box plot is constructed using the empirical distribution of the SNR improvements $10\mathrm{log}_{10}(\mathrm{SNR}_{\mathrm{DPC}}/\mathrm{SNR}_{\mathrm{sw}})$ and $10\mathrm{log}_{10}(\mathrm{SNR}_{\mathrm{DPC}}/\mathrm{SNR}_{\mathrm{dp}})$. We observe from Fig. 5 that the median SNR improvement with DPC phased arrays is about $1.9$ dB over the switched-polarization architecture and about $0.4$ dB over the dual-polarization architecture. The poor performance with the switched-polarization architecture is due to the fact that it performs beamforming with just one set of dipole antennas.
\begin{figure}[h!]
\centering
\subfloat[$10\mathrm{log}_{10}(\mathrm{SNR}_{\mathrm{DPC}}/\mathrm{SNR}_{\mathrm{sw}})$]{\includegraphics[trim=2cm 6.5cm 2.5cm 7.5cm,clip=true,width=4.45cm, height=4.6cm]{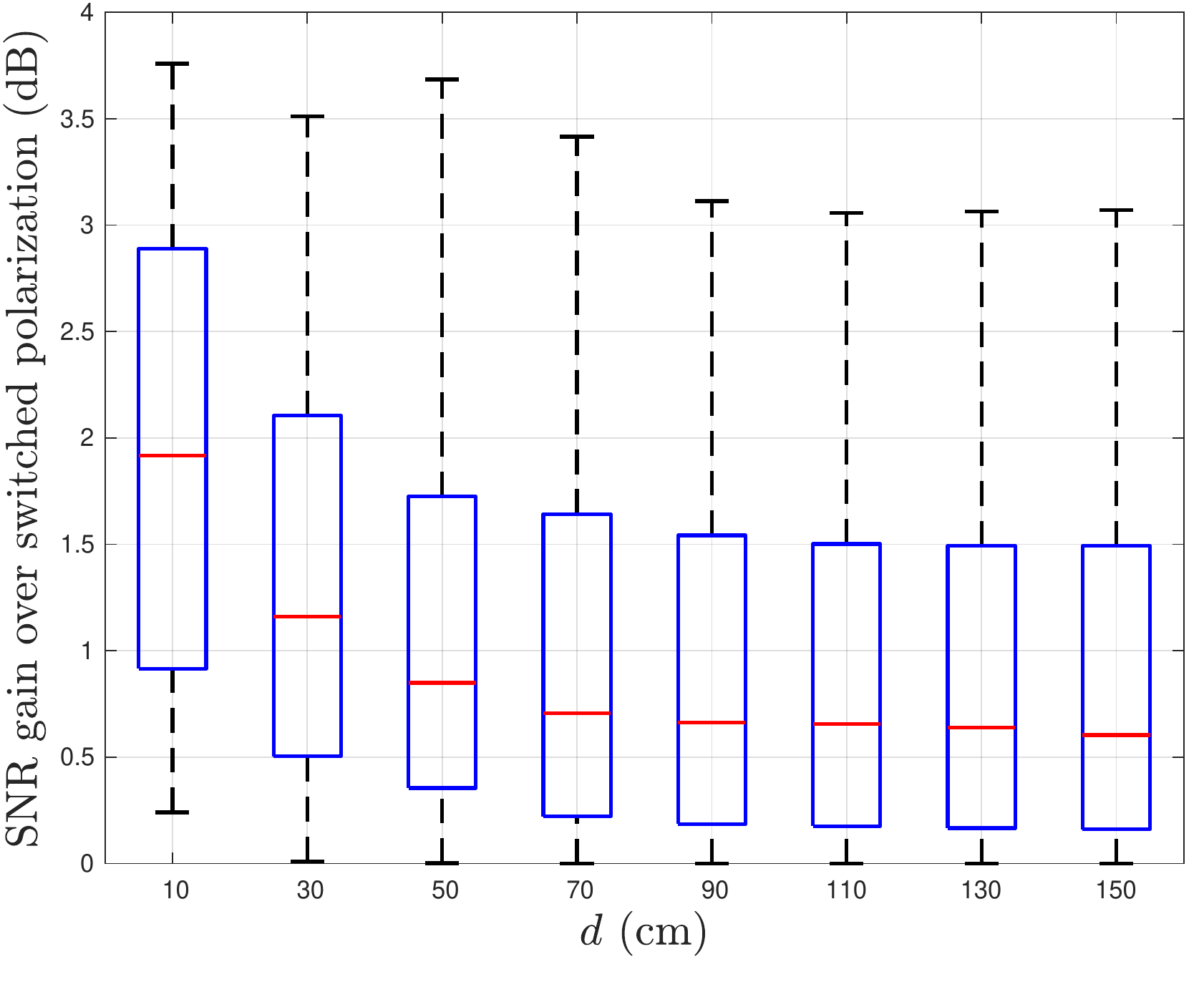}\label{fig:SNRimp_SW}} 
\subfloat[$10\mathrm{log}_{10}(\mathrm{SNR}_{\mathrm{DPC}}/\mathrm{SNR}_{\mathrm{dp}})$]{\includegraphics[trim=2cm 6.5cm 2.5cm 7.5cm,clip=true,width=4.45cm, height=4.6cm]{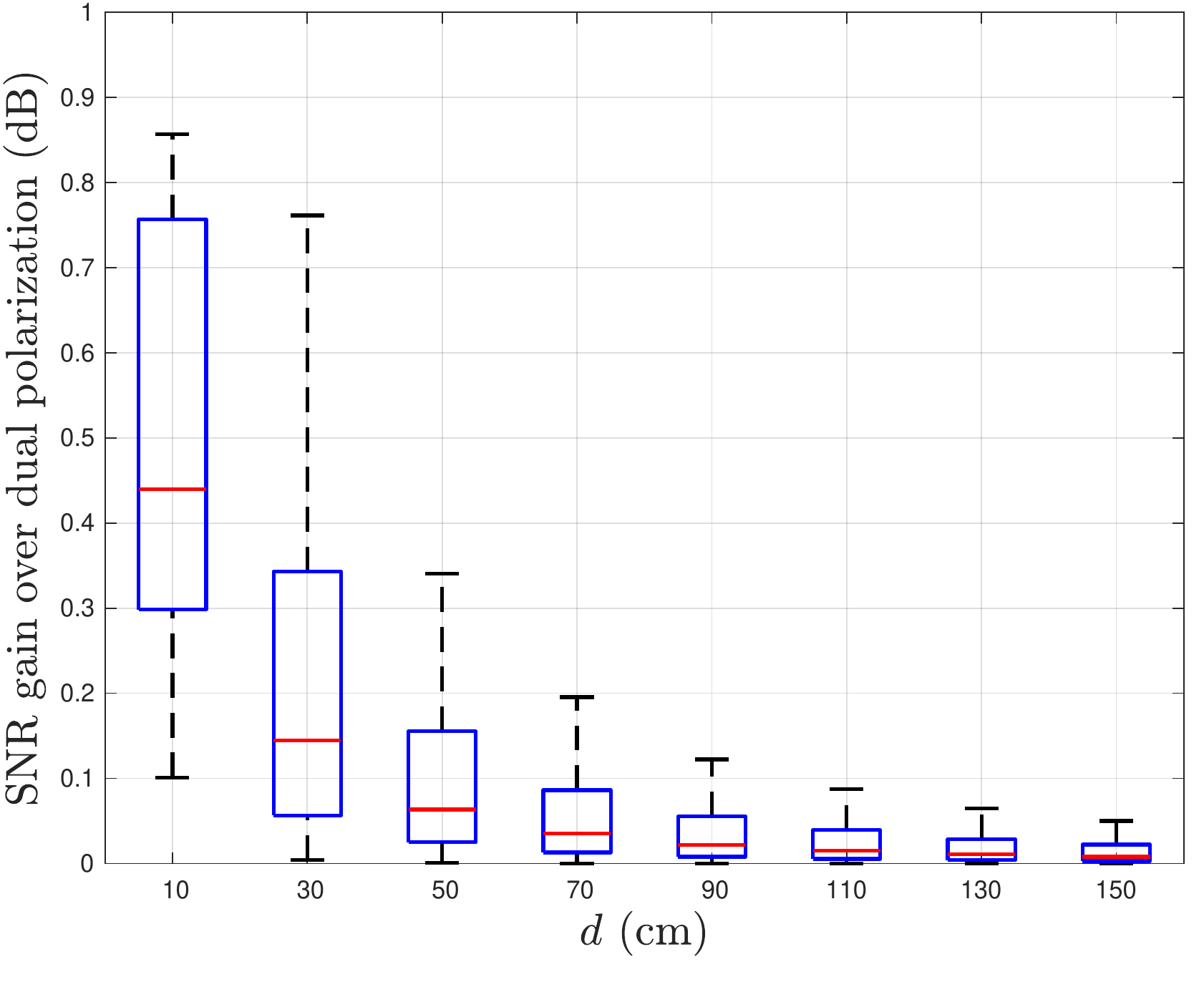}\label{fig:SNRimp_Dual}}
\caption{\small SNR improvement with the use of a DPC phased array over switched- or dual- polarization architectures.  The SNR improvement is plot as a function of the transceiver distance $d$ for $\alpha=30^{\circ}$.
\normalsize}
\end{figure}
\begin{figure}[h]
\centering
\includegraphics[trim=2cm 7cm 2.5cm 7.75cm,width=0.39\textwidth]{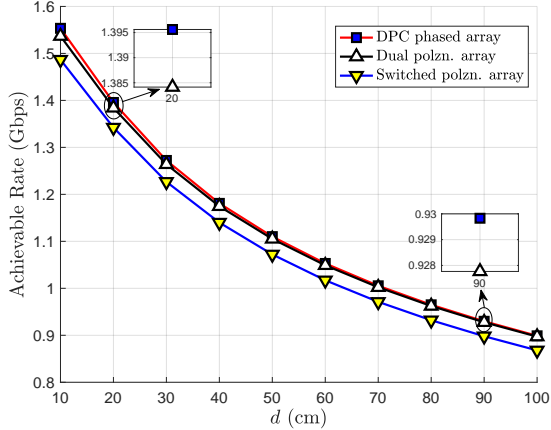}
\caption{\small The achievable rates when DPC phased array,  switched- and dual- polarization architectures are used at the TX. For each distance $d$, the rate is averaged over several orientations of the RX.}
\normalsize
  \label{fig:Rate_TX_RX_dist}
\end{figure}
\par Now, we study the performance with DPC as a function of the transceiver distance $d$. Here, we assume that the center of the RX is located along a ray that makes $\alpha=30^{\circ}$ with the $\hat{\mathbf{z}}$.  For every location of the RX center determined by $d$,  the receive dipole is rotated about its axis in steps of $10^{\circ}$ along azimuth and elevation.  Then,  a distribution of the received SNR is constructed with these channel realizations for every $d$.  We observe from Fig. 6 that the dual-polarization array performs as good as the DPC array, at larger transceiver distances. This is because the optimal polarization configuration at the TX becomes spatially invariant in the far field regime, and such a configuration can be realized with the common dual-polarization array. 
\par We now plot the ergodic achievable rate with the transceiver distance $d$.  The rate is obtained by averaging $B\mathrm{log}_2(1+\mathrm{SNR})$ across all the channel realizations obtained by rotating the RX., i.e., by changing $\hat{\mathbf{v}}$.  We use $\Ptx=1\, \mathrm{mW}$ in Fig. \ref{fig:Rate_TX_RX_dist}. We notice that the achievable rate with DPC is almost the same as that with dual-polarization, as an SNR improvement of $0.4\, \mathrm{dB}$ results in a negligible increase in the rate. 
\section{Conclusions and future work}
In this paper, we showed that the optimal polarization configuration varies spatially in near-field communication unlike the far-field case.  Specifically, we observed that the polarization angle varies across the antenna array.  We studied how DPC phased arrays achieve the desired configuration, thereby resulting in a higher received SNR than switched- and dual-polarization phased arrays.  Our simulations indicate that DPC phased arrays are useful over conventional architectures only at short distances.  In future, we will investigate the trade-off between power consumption and the achievable rate for the DPC phased array.  
\bibliographystyle{IEEEtran}
\bibliography{refs}

\begin{thebibliography}{10}
\providecommand{\url}[1]{#1}
\csname url@samestyle\endcsname
\providecommand{\newblock}{\relax}
\providecommand{\bibinfo}[2]{#2}
\providecommand{\BIBentrySTDinterwordspacing}{\spaceskip=0pt\relax}
\providecommand{\BIBentryALTinterwordstretchfactor}{4}
\providecommand{\BIBentryALTinterwordspacing}{\spaceskip=\fontdimen2\font plus
\BIBentryALTinterwordstretchfactor\fontdimen3\font minus
  \fontdimen4\font\relax}
\providecommand{\BIBforeignlanguage}[2]{{%
\expandafter\ifx\csname l@#1\endcsname\relax
\typeout{** WARNING: IEEEtran.bst: No hyphenation pattern has been}%
\typeout{** loaded for the language `#1'. Using the pattern for}%
\typeout{** the default language instead.}%
\else
\language=\csname l@#1\endcsname
\fi
#2}}
\providecommand{\BIBdecl}{\relax}
\BIBdecl

\bibitem{NF_fraunhoffer}
T.~S. Rappaport \emph{et~al.}, \emph{Wireless communications: \text{Principles}
  and practice}.\hskip 1em plus 0.5em minus 0.4em\relax Prentice hall PTR New
  Jersey, 1996, vol.~2.

\bibitem{THz_applications}
V.~Petrov, T.~Kurner, and I.~Hosako, ``\text{IEEE 802.15. 3d}: First
  standardization efforts for sub-terahertz band communications toward
  \text{6G},'' \emph{IEEE Commun. Mag.}, vol.~58, no.~11, pp. 28--33, 2020.

\bibitem{NF_focus}
D.~Headland, Y.~Monnai, D.~Abbott, C.~Fumeaux, and W.~Withayachumnankul,
  ``Tutorial: Terahertz beamforming, from concepts to realizations,''
  \emph{Appl. Photonics}, vol.~3, no.~5, p. 051101, 2018.

\bibitem{sw_dual_pol}
O.~Jo, J.-J. Kim, J.~Yoon, D.~Choi, and W.~Hong, ``Exploitation of
  dual-polarization diversity for \text{5G} millimeter-wave \text{MIMO}
  beamforming systems,'' \emph{IEEE Trans. on Ant. and Prop.}, vol.~65, no.~12,
  pp. 6646--6655, 2017.

\bibitem{dual_pol_BF2}
J.~Song, J.~Choi, S.~G. Larew, D.~J. Love, T.~A. Thomas, and A.~A. Ghosh,
  ``Adaptive millimeter wave beam alignment for dual-polarized \text{MIMO}
  systems,'' \emph{IEEE Trans. on Wireless Commun.}, vol.~14, no.~11, pp.
  6283--6296, 2015.

\bibitem{DPC_hw2}
A.~Safaripour, S.~M. Bowers, K.~Dasgupta, and A.~Hajimiri, ``Dynamic
  polarization control of two-dimensional integrated phased arrays,''
  \emph{IEEE Trans. on Microwave Theory and Tech.}, vol.~64, no.~4, pp.
  1066--1077, 2016.

\bibitem{NF_rotULA}
H.~Do, N.~Lee, and A.~Lozano, ``Reconfigurable \text{ULAs} for line-of-sight
  \text{MIMO} transmission,'' \emph{IEEE Trans. on Wireless Commun.}, vol.~20,
  no.~5, pp. 2933--2947, 2020.

\bibitem{NF_chest}
N.~J. Myers, J.~Kaleva, A.~T{\"o}lli, and R.~W. Heath, ``Message passing-based
  link configuration in short range millimeter wave systems,'' \emph{IEEE
  Trans. on Commun.}, vol.~68, no.~6, pp. 3465--3479, 2020.

\bibitem{NF_chest2}
M.~Cui and L.~Dai, ``Channel estimation for extremely large-scale \text{MIMO}:
  Far-field or near-field?'' \emph{IEEE Trans. on Commun.}, 2022.

\bibitem{InFocus}
N.~J. Myers and R.~W. Heath, ``Infocus: A spatial coding technique to mitigate
  misfocus in near-field \text{LoS} beamforming,'' \emph{IEEE Trans. on
  Wireless Commun.}, 2021.

\bibitem{WB_BF}
M.~Cui, L.~Dai, R.~Schober, and L.~Hanzo, ``Near-field wideband beamforming for
  extremely large antenna array,'' \emph{arXiv preprint arXiv:2109.10054},
  2021.

\bibitem{JBFPT}
S.~Sugiura, Y.~Kawai, T.~Matsui, T.~Lee, and H.~Iizuka, ``Joint beam and
  polarization forming of intelligent reflecting surfaces for wireless
  communications,'' \emph{IEEE Trans. on Vehicular Tech.}, vol.~70, no.~2, pp.
  1648--1657, 2021.

\bibitem{HW_pol1}
D.~Blanco, J.~L. G{\'o}mez-Tornero, E.~Rajo-Iglesias, and N.~Llombart,
  ``Radially polarized annular-slot leaky-wave antenna for three-dimensional
  near-field microwave focusing,'' \emph{IEEE Ant. and Wireless Prop. Lett.},
  vol.~13, pp. 583--586, 2014.

\bibitem{HW_pol2}
A.~J. Martinez-Ros, J.~L. G{\'o}mez-Tornero, V.~Losada, F.~Mesa, and F.~Medina,
  ``Non-uniform sinusoidally modulated half-mode leaky-wave lines for
  near-field focusing pattern synthesis,'' \emph{IEEE Trans. on Ant. and
  Prop.}, vol.~63, no.~3, pp. 1022--1031, 2014.

\bibitem{HW_pol3}
A.~Sharma, I.~J.~G. Zuazola, R.~Martinez, J.~C. Batchelor, A.~Perallos, and
  L.~d.-H. Ariet, ``Optimal \text{E}-field vector combination for a highly
  focused antenna-array,'' \emph{IEEE Ant. and Wireless Prop. Lett.}

\bibitem{IOT_dipole}
L.~Chen, W.~Hu, K.~Jamieson, X.~Chen, D.~Fang, and J.~Gummeson, ``Pushing the
  physical limits of \text{IoT} devices with programmable metasurfaces,'' in
  \emph{Proc. of the 18th USENIX Symp. on Networked Sys. Des. and Implem. (NSDI
  21)}, 2021, pp. 425--438.

\bibitem{Ant_text}
C.~A. Balanis, \emph{Antenna theory: \text{Analysis} and design}.\hskip 1em
  plus 0.5em minus 0.4em\relax John wiley \& sons, 2015.

\bibitem{MassMIMO}
P.~Chandhar, D.~Danev, and E.~G. Larsson, ``Massive \text{MIMO} for
  communications with drone swarms,'' \emph{IEEE Trans. on Wireless Commun.},
  vol.~17, no.~3, pp. 1604--1629, 2017.

\end{thebibliography}
\end{document}